**Revision of the linear stability paradox for known bounded shear flows**

Sergey G. Chefranov [1);*)] and Alexander G. Chefranov [2)]

[1)] A. M. Obukhov Institute of Atmospheric Physics, Moscow, Russia

[*)] The author for correspondence: schefranov@mail.ru ;

schefranov@candqrc.cn

[2)] Eastern Mediterranean University, Famagusta, Northern Cyprus

Alexander.chefranov@emu.edu.tr

## Abstract

The well-known paradox of linear stability for the Hagen-Poiseuille (HP) and Plane Couette (PC) flows is not solved up to now and is bypassed on the basis of the non-linear mechanisms consideration. We prove that it is arising only due to an idealized assumption of an exact space periodicity for the small hydrodynamic perturbations. When finite non-zero viscosity is taken into account only quasi-periodic in space perturbations can be considered in the frame of linear stability theory. For the quasi-periodic in longitudinal direction disturbances the linear instability of the HP flow, Plane Poiseuille flow (PP) and PC flow at the finite Reynolds numbers, is obtained. The generalization of Landau's critical velocity for the vortexes arising in the laminar HP, PP and PC flows of classical fluids also stated.

## 1. Introduction

In the theory of turbulence, the mechanism of transition from a laminar vortex-free stationary flow regime to a turbulent regime remains unclear [1]-[16]. Indeed, for a number of limited shear stationary flows there is still no correspondence between the conclusions of the linear theory of hydrodynamic stability and the data of experimental observations. [1]-[3], [16]. This primarily applies to the Hagen-Poiseuille flow (HP), as well as to the plane Couette flow (PC).

For these shear flows, linear theory establishes the absence of exponential instability with respect to extremely small amplitude disturbances for any arbitrarily large Reynolds numbers (see [2], [3],[16] and the links provided there).

Before the appearance of the famous article by Heisenberg (1924) [17], the results of which were not immediately recognized [3], the plane Poiseuille flow (PP) was also included in this group of shear currents. Moreover, for a plane Poiseuille flow, Lyapunov stability is known with respect to finite amplitude perturbations in the case of a flow of an ideal medium with zero viscosity [3]. The reason for the long-term non-recognition of the work [17] is related to the dominant and still prevailing idea that the effect of viscosity always contributes only to the stabilization of the system. That is why the result obtained by Heisenberg [17] for the flow of a viscous medium has not been recognized for more than twenty years [3].

However, the conditions of linear instability of the PP flow obtained in [17] and even their refinement in [18] still turn out to be several times higher than the threshold values of the Reynolds number observed in the experiment [2], [3].

In this paper, an explanation of this quantitative difference is given, which is linked to the traditionally used idealized [19] assumption of a strict longitudinal periodicity of small disturbances along the direction of movement of the main stream. This idealization, as shown below, turns out to be completely unacceptable for systems in which it is possible to realize dissipative instability due to the presence of low, but nonzero viscosity.

Indeed, by now many examples of dissipative instability have been known, when the influence of low but nonzero viscosity causes exponentially fast realization of linear instability in open systems [20]-[32].

It should be noted that long before Heisenberg's work [17], it was shown that taking into account even a small frictional force leads to instability of the mechanical system [20]. In [20](see also [21], [28]- [30]). It is shown that only with a nonzero frictional force proportional to velocity, a two-dimensional oscillator with its own frequency $\omega_0$ in a coordinate system rotating with frequency $\Omega$ has an unstable zero stationary state in the case of sufficiently fast rotation when $\Omega > \omega_0$.

However, this classic example of dissipative instability is still not widely known. For example, there is no mention of him even in such fundamental works as [33] and [34]. Although [33] and [34] consider a similar problem of a two-dimensional linear oscillator (Foucault's pendulum) in a rotating coordinate system, but without taking into account friction and only with a relatively slow rotation of the system compared with the natural frequency of the linear oscillator. At the same time, it was shown only in relatively recent works [28],[29] that this dissipative instability mechanism also leads to a violation of chiral symmetry, which was not noticed in [20] and [21]. As a result, it was found that it is the dissipative mechanism of instability that can provide an explanation for the vortex cyclone-anticyclone asymmetry observed on the fast rotating planets [28], [29]. This effect also turns out to be important for understanding the mechanism of formation of the initial stage of the origin of tropical cyclones [30].

Nevertheless, the very fact that the instability of the PP flow is caused precisely by the action of nonzero viscosity is well known and even specifically noted in [3] and [35]. However, an explicit analysis of the dissipative mechanism of linear instability of this flow, as well as for the HP flow and PC flow was not considered at all until recently.

Instead, to circumvent this problem of linear theory, only various options for considering finite amplitude perturbations were used. [2], [3], [7]- [16], in which viscosity, on the contrary, plays only a stabilizing role, for example, in the processes of transient growth [36]-[38].

In particular, [11] also used the rejection of the traditional, so-called normal form of perturbations. [2], [3], [35]. However, this allowed only the rejection of the modality of small perturbations while maintaining the assumption of their strict longitudinal periodicity, which clearly idealizes the observed variability of real perturbations [19].

As a result, [11] considered only non-modal perturbations with algebraic (power-law) growth of small perturbations, which were introduced earlier in [36]-[38] to circumvent the problem of linear stability of HP and PC flows.

However, these methods of circumventing the problem of linear stability of limited shear flows do not sufficiently clarify the mechanism of transition to a turbulent regime, as the linear theory of stability that provides in many other cases [16]. Indeed, linear theory usually provides a completely acceptable estimate of the boundary of the instability region, which is only slightly refined when taking into account the nonlinear effects associated with the finiteness of the amplitude of real disturbances. Therefore, the finite-amplitude transition mechanisms proposed so far cannot replace the direct elimination of the paradox of linear stability in the HP and PC flows.

Moreover, according to experimental data [39], for the loss of stability of the laminar flow of HP at a fixed amplitude of disturbances, the main role is played not by the magnitude of the amplitude of the disturbance, but by the frequency response of disturbances having an almost periodic character on the eve of the change of the laminar flow regime to the turbulent regime.

The noted results of [39], as well as the rejection of the idealization mentioned above and in [19] in the form of strict periodicity of perturbations, were taken into account in [40]-[43].  This has already made it possible not to circumvent, but directly eliminate the paradox of linear stability and obtain conditions for linear instability of the HP, PP and PC flows. At these conditions, corresponding to finite threshold Reynolds numbers, are fulfilled, it is precisely by taking into account the finite nonzero viscosity that exponentially fast rather than power-law algebraic growth of perturbations over time is realized.

As a result, in [40]-[43], the so-called conditionally periodic small perturbations were considered instead of longitudinally periodic perturbations [33].  In this case, the evolution of perturbations is characterized not only by the Reynolds number, but also by the ratio of the longitudinal periods characterizing various transverse modes determined by the type of boundary conditions. For example, for the flow of HP in the case of two radial modes in [40], [41], a minimum threshold Reynolds number Re=448 was obtained, which is consistent with the threshold Reynolds number Re=420, characteristic of the condition for the occurrence of Tollmien-Schlichting waves observed in the boundary layer.

This finite value of the Reynolds threshold number is obtained when the value of the additional parameter p=1.527 characterizes the ratio of the longitudinal periods of the radial modes corresponding to the first and second zero of the first-order Bessel function. On the contrary, for values of the parameter p equal to an integer (or the inverse of an integer), the Reynolds threshold number is no longer finite, but tends to infinity. That corresponds to the conclusion of linear theory, which is still known, based on the use of strictly longitudinally periodic small perturbations with the same period for all transverse radial mods.

A similar consideration in the framework of linear stability theory, but for the flow of PP in [43] led to an estimate of the threshold Reynolds number $\mathrm{Re}_{th} \approx 1035$ (for the ratio $p \approx 0.49$ of the periods of two transverse modes), which is already in good agreement with the value $\mathrm{Re}_{\exp} \approx 1080$, obtained experimentally in [44] (see also [3], [45]-[47]). In [18], the normal form of small perturbations is used with the same longitudinal period for all transverse modes (that is, with the parameter value $p = 1$ in the case of two modes). As a result, [18] obtained the best known estimate of the Reynolds threshold number $\mathrm{Re}_{th} \approx 5772$ for the linear instability of the plane Poiseuille flow, which, however, is more than five times higher than the above experimentally observed value.

In [43] and for the Couette plane flow, a threshold value $\mathrm{Re}_{th} \approx 305$ of the Reynolds number was obtained in the framework of linear theory (with the ratio $p \approx 0.746$ of the longitudinal periods of two transverse modes), which also agrees well with the threshold value $\mathrm{Re}_{\exp} = 325 \pm 5$ observed in the experiment [48]-[50].

Thus, threshold values of the Reynolds number consistent with experiment have already been obtained in [40]-[43], which eliminates the paradox of linear stability of the shear flows under consideration due to the rejection of the normal form of small disturbances, that assuming the same longitudinal periodicity for different transverse modes. This result was obtained in [40], [41] using the approximate Bubnov-Galerkin method not only for the radial, but also in longitudinal coordinates. In [42] and [43], instead, another also approximate (energy) method is used, which is based on the need to average over the longitudinal period of one of the modes. In all these cases, strict longitudinal periodicity is still assumed for each of the modes, although the values of the periods corresponding to different transverse modes may differ from each other, unlike the normal form of small perturbations.

In our present paper, a more accurate study of the evolution of radial perturbation modes is carried out, when there is no longer a need to use approximate methods to describe their longitudinal variability. In particular, unlike [40]-[43], strict longitudinal periodicity is not assumed even for each of the radial disturbance modes, which is also more consistent with experimental observations [19]. Instead, to account for the longitudinal spatial variability of each of the linearly interacting radial modes of HP flow disturbances, almost periodic boundary conditions along the longitudinal coordinate related to the so-called regular boundary conditions [51]-[53] are considered (see also [54], [55]).

In the next section 2, the linear theory of stability of the HP flow with respect to extremely small amplitude perturbations of the tangential component of the velocity field is considered, provided that the perturbations are axially symmetric (there is no dependence on the angular variable for the velocity field and pressure of small perturbations). The axial symmetry condition simplifies the consideration as much as possible, since it reduces to a closed description of only the tangential component of the perturbation velocity field at zero values of small perturbations of other velocity and pressure components. In section 3, a similar theory is developed for the plane Poiseuille flow (PP) and the plane Couette flow (PC), and the conditions of linear instability corresponding to these flows are obtained with respect to quasi-periodic disturbances of the transverse component of the velocity field that are extremely small in amplitude.

In section 4, a generalization of the Landau theory [22] is obtained, which also gives new instability conditions of the HP, PP and PC flows of the classical fluid.

## 2. Linear instability of the Hagen-Poiseuille (HP) flow

1. For simplicity, let us consider the evolution of extremely small disturbances only for the tangential component of velocity field $u_\varphi \neq 0$ to the Hagen-Poiseuille (HP) steady flow in cylindrical coordinates $z; r; \varphi$:

$$\frac{\partial u_\varphi}{\partial t} + U_z(r)\frac{\partial u_\varphi}{\partial z} = \nu\left(\frac{\partial^2 u_\varphi}{\partial z^2} + \frac{1}{r}\frac{\partial}{\partial r}r\frac{\partial u_\varphi}{\partial r}\right);$$

$$U_z(r) = U_{\max}\left(1 - \frac{r^2}{R^2}\right); U_r = U_z = 0: HP - flow \qquad (1)$$

$$u_r = u_z = 0; \frac{\partial p}{\partial \varphi} = \frac{\partial u_\varphi}{\partial \varphi} = 0$$

The boundary conditions for solution of Eq.(1) has representation:

$$u_\varphi(r = R) = u_\varphi(r = 0) = 0 \qquad (2)$$

The solution of the Eq. (1) at the boundary conditions (2) may be considered in the form:

$$u_\varphi(z;r;t) = U_{\max} \sum_{n=1}^{N} A_n(x;\tau) J_1(j_{1,n} y);$$

$$x = z/R; \quad y = r/R; \quad \tau = t\nu/R^2$$

(3)

In (3) $J_1$ - is the Bessel function of the first order where $J_1(j_{1,n}) = 0; n = 1,2,...,N$.

The perturbation of the tangential component of the velocity field in (3) corresponds to the following representations for the components of the vortex field perturbation:

$$\omega_z(x;y;\tau) = \frac{1}{r}\frac{\partial(ru_\varphi)}{\partial r} = \frac{U_{\max}}{R} \sum_{n=1}^{N} A_n(x;\tau) j_{1,n} J_0(j_{1,n} y);$$

$$\omega_r(x;y;\tau) = -\frac{\partial u_\varphi}{\partial z} = -\frac{U_{\max}}{R} \sum_{n=1}^{N} \frac{\partial A_n(x;\tau)}{\partial x} J_1(j_{1,n} y);$$

(4)

$$\omega_\varphi = 0$$

After introduction of the representation (3) into Eq. (1) and by using the Galerkin approximation on the radial coordinate, it is possible to obtain in the dimensionless form the next system of equations to the amplitudes of various radial modes $A_m$ (see also (3) in [41]):

$$\frac{\partial A_m}{\partial \tau} + j_{1,m}^2 A_m - \frac{\partial^2 A_m}{\partial x^2} + \text{Re} \sum_{n=1}^{N} P_{nm} \frac{\partial A_n}{\partial x} = 0$$

$$m = 1,2,...N; -\infty < x < \infty; \text{Re} = \frac{U_{\max} R}{\nu}$$

(5)

$$P_{nm} = \frac{2}{J_2^2(j_{1,m})} \int_0^1 dy\, y(1-y^2) J_1(j_{1,n} y) J_1(j_{1,m} y)$$

(6)

In (5), (6) $J_1, J_2$ - are the Bessel functions of the first and second order, respectively, $\text{Re}$ - is the dimensionless Reynolds number; $U_{\max}; R; \nu$ -are the maximum flow velocity of HP on the pipe axis, the pipe radius, and the kinematic viscosity coefficient.

As already noted in the Introduction, in [40]-[42], based on the consideration of the system of equations (5) for the case of two linearly interacting radial modes (when in (5) N=2) having different values of the periods of longitudinal variability along the pipe axis. In these works, conditions for the linear instability of the HP flow at finite Reynolds numbers are obtained. However, this is possible only for cases when the ratio of these two different longitudinal periods is not equal to one of the following three possible values $p = p_k ; p = p_{1/k} ; p = p_{\sqrt{k}}$, for which the threshold Reynolds number $\mathrm{Re}_{th} \to \infty$ ($\mathrm{Re}^{-1}_{th}(p) \cong \sin(\pi p)\sin(\pi / p)\sin(\pi(p+1/p))$) is infinitely large:

$$
\begin{aligned}
&p_k = k, k = 1,2,...; \\
&p_{1/k} = 1/k; k = 1,2,...; \\
&p^{\pm}_{\sqrt{k}} = \frac{k+1 \pm \sqrt{(k+1)^2 - 4}}{2}, k = 1,2,..
\end{aligned}
\tag{7}
$$

Thus, according to (7), linear instability of the HP flow is impossible even in some cases of incommensurable longitudinal periods, when the ratio $p^{\pm}_{\sqrt{k}}$ of periods is equal to an irrational number. For example, in the case of $k = 2$ a value $p^{+}_{\sqrt{2}} = 1 + G; G = (1 + \sqrt{5})/2 \approx 1.618..$, where $G$ - is an irrational number corresponding to the known golden ratio. In particular, in [40], [41], with a ratio $p = 1.53$ of longitudinal periods, the HP flow becomes unstable relative to extremely small amplitude disturbances when the threshold Reynolds number $\mathrm{Re}_{th} \approx 448$ is exceeded.  This value is close to the threshold Reynolds number 420, which characterizes the condition for the occurrence of the observed Tolmin-Schlichting waves in the boundary layer, which is also caused by the mechanism of dissipative instability.

 However, as the number of radial modes under consideration increases, the threshold Reynolds number increases, and for 100 radial modes, the threshold number of linear instability of the HP flow already exceeds 600 [42]. In the case of one radial mode having strictly periodic variability along the pipe axis, linear instability of the HP flow also turns out to be impossible, as in the case of two radial modes having a ratio of longitudinal periods equal to one of the values (7).

We show that even in the case of a single radial mode in (5), linear instability of the HP flow still turns out to be possible. It is so if we take into account the influence of the viscous dissipation effect, which leads to the need to consider replacing the strictly periodic variability of disturbances along the pipe axis with an almost periodic behavior when the longitudinal coordinate changes.

Consider system (5) for the case of a single nonzero mode with an arbitrary exponent m (m=1, 2, 3, .. ), that is, for the case N=1 in (5), under the boundary condition characterizing the almost periodic longitudinal variability of this radial mode [9]-[11]:

$$\beta A_m(x=0;\tau) = A_m(x=1/k_m;\tau);$$

$$\beta\left(\frac{\partial A_m}{\partial x}\right)_{x=0} = \left(\frac{\partial A_m}{\partial x}\right)_{x=1/k_m} \tag{8}$$

$$1-\beta << 1$$

Boundary conditions of this type are called regular in [11] and they are used in this form in [10], and their modification is considered in [9].

Only at a strictly zero viscosity value does the coefficient value $\beta$ tend to unity and condition (8) turns into a condition of strict longitudinal periodicity of radial modes.

For example, a vivid illustration of such a mechanism of action of viscosity is an example of replacing a strictly time-periodic mode change $a(t) = a_0 \cos\omega t$ in the case of zero viscosity on the dependence $a(t) = a_0 e^{-\nu k^2 t} \cos\omega t$ at a finite viscosity value $\nu \neq 0$.

The analogue of the relation (8) in this case will be the condition $\tilde{\beta}a(t=0) = a(t=2\pi/\omega)$, where $\tilde{\beta} = \exp(-\nu k^2 2\pi/\omega)$. This condition replaces the strict periodicity condition $a(t=0) = a(2\pi/\omega)$, which is valid only in the case of zero viscosity when $\tilde{\beta} = 1$.

For the case of one non-zero mode with an arbitrary number $m$ in (5), we obtain the equation for this mode in the form:

$$\frac{\partial A_m}{\partial \tau} + j_{1,m}^2 A_m + \operatorname{Re} P_{mm} \frac{\partial A_m}{\partial x} = \frac{\partial^2 A_m}{\partial x^2} \tag{9}$$

 We will look for a solution to equation (9) under almost periodic boundary conditions (8) in the form $A_m(\tau;x) = e^{S\tau}A_{m0}(x)$, when the partial differential equation (9) reduces to the equation in ordinary derivatives:

$$\frac{d^2 A_{m0}}{dx^2} - \operatorname{Re} P_{mm}\frac{dA_{m0}}{dx} - S - j_{1,m}^2 = 0; \qquad (10)$$

Let us find a solution to the equation (10) under the boundary condition (8) in the form:

$$A_{m0} = C_- \exp(-x\lambda_-) + C_+ \exp(x\lambda_+);$$
$$\lambda_\pm = \frac{\sqrt{\operatorname{Re}^2 P_{mm}^2 + 4(S + j_{1,m}^2)} \pm \operatorname{Re} P_{mm}}{2} \qquad (11)$$

From the boundary conditions (8) and the finiteness condition of the solution (11), for arbitrary integration constants $C_\pm \neq 0$, we obtain the following dispersion relation:

$$\left(\lambda_+ + \lambda_-\right)\left(\beta - \exp\left(-\frac{\lambda_-}{k_m}\right)\right)\left(\beta - \exp\left(\frac{\lambda_+}{k_m}\right)\right) = 0 \qquad (12)$$

In particular, from (12) and the condition $\beta = \exp(-\lambda_-/k_m)$, taking into account the limit $0 < 1 - \beta \ll 1$ (at which an inequality holds $\operatorname{Re} P_{mm} \gg k_m \ln(1/\beta)$), we obtain the following dispersion equation for determining the increment value $S$:

$$S = -j_{1,m}^2 + \operatorname{Re} P_{mm} k_m \ln\left(\frac{1}{\beta}\right) \qquad (13)$$

 Thus, according to representation (13), linear instability of the HP flow with exponential growth of extremely small initial amplitude perturbations of the tangential component of the velocity field turns out to be possible for any finite Reynolds numbers exceeding the following threshold value:

$$\operatorname{Re} > \frac{j_{1,m}^2}{P_{mm} k_m \ln\left(\frac{1}{\beta}\right)} > \frac{j_{1,1}^2}{P_{11} k_1 \ln\left(\frac{1}{\beta}\right)} \approx \frac{20.89}{k_1 \ln\left(\frac{1}{\beta}\right)} \qquad (14)$$

Where in (14) $j_{1,1} \approx 3.73 < j_{1,2} \approx 7.01; P_{11} \approx P_{22} \approx 0.666.$

For the solution of the dispersion equation (12) considered in (13) and (14), corresponding to the first radial mode $A_{10}(x)J_1(j_{1,1}y)$, the dependence of the perturbation of the tangential component of the velocity field on the longitudinal coordinate according to (11) has the form $u_\varphi \propto e^{S\tau}\exp(-xk_1\ln(1/\beta)) \propto e^{S\tau}\exp(-x/l_\nu)$. Here, the dimensionless parameter $l_\nu = k_\nu^{-1}; k_\nu \propto k_1\ln(1/\beta)$ characterizes the scale of the longitudinal variability of the disturbance field. To determine the relationship between this scale and the viscosity value, we use estimates of the characteristic transverse dimensions $\varepsilon$ of the vortex ring with radius $R$, which arise precisely due to non-zero viscosity during the interaction of the flow in the pipe with a rigid boundary $\varepsilon/R \propto k_1\ln(1/\beta) \propto \mathrm{Re}^{-1/2}$ [3], [56], [57].

Indeed, for the Hagen-Poiseuille flow, the characteristic length scale $\varepsilon$, due to the action of viscosity has estimation $\varepsilon \approx \sqrt{\nu(dU_z/dr)_{r=R}^{-1}} \propto R(2\mathrm{Re})^{-1/2}$ and determines in (13) the type of dependence $\beta \approx 1 - \varepsilon/R + O(\varepsilon^2/R^2); \varepsilon/R \ll 1$ on the Reynolds number [57]:

$$k_1\ln 1/\beta \approx \varepsilon/R = (2\mathrm{Re})^{-1/2}; \mathrm{Re} \gg 1 \qquad (15)$$

According to (15) and (14), linear instability of the HP flow can occur as a result of the generation of annular vortices near the walls of the pipe at threshold values of the Reynolds number:

$$\mathrm{Re} > \mathrm{Re}_{1HP} \approx 872{,}78 \qquad (16)$$

In the case of a radial mode with the number m=2, under the condition of linear instability of the flow, in (14) it is need to replaced $j_{1,1}^2 \to j_{1,2}^2 \approx 49.14; P_{11} \to P_{22} \approx 0.666$. In that case an exponential instability of HP flow takes place for the larger threshold Reynolds numbers:

$$\mathrm{Re} > \mathrm{Re}_{2HP} \approx 1088814 \qquad (17)$$

The obtained value of the threshold Reynolds number in (16) is approximately twice as high as the threshold Reynolds number $\mathrm{Re}_{TS} \approx 420$, for the occurrence of Tollmien-Schlichting waves in the boundary layer [58]-[60]. As noted above, the minimum threshold Reynolds number obtained in [40], [41] for the HP flow when considering in (5) two radial modes (N=2) with different longitudinal periods is close to this value and is equal to $\mathrm{Re}_{th} \approx 448$.

Note that with an increase in the number of radial modes in (5), the obtained minimum value of the threshold Reynolds number already tends to increase noticeably, and in [42], when considering one hundred radial modes (N=100) in (5), this value is equal to $\mathrm{Re}_{th} \approx 600$ .

### 3. Linear instability of the Plane Couette (PC) and Plane Poiseuille (PP) flows

Let us consider the evolution of an extremely small amplitude disturbance of the stationary velocity field of a plane Poiseuille flow (PP) and a plane Couette flow (PC) of a viscous incompressible fluid.

In the Cartesian coordinate system $x, y, z$ , we will describe the flows of PP and PC in a dimensionless form, choosing the positive direction of the axis $x$ in the direction of fluid movement when two rigid flat boundaries bounding these flows are located at a distance $H$ from each other along the axis $z$ (see Fig.1).

In a fixed coordinate system, the center of which is equidistant from the rigid boundaries, for dimensionless representations of coordinates through dimensional coordinates, we use the representation $x = 2x'/H, y = 2y'/H, z = 2z'/H$ . By analogy with the consideration of the problem of linear stability of the HP flow, we limit ourselves to studying the evolution of only one transverse component of the velocity field disturbance $u_y(x; z; \tau = 4t\nu / H^2)$ , which does not depend on the coordinate $y$ in the absence of a pressure gradient disturbance in this direction.

The evolution of the disturbance can be described using the following equation for the velocity field $u_y(\tau; x; z)$ , assuming that the remaining components of the velocity field disturbance are zero $u_x = u_z = 0$ :

$$\frac{\partial u_y}{\partial \tau} + U_x(z)\,\mathrm{Re}\,\frac{\partial u_y}{\partial x} = \frac{\partial^2 u_y}{\partial x^2} + \frac{\partial^2 u_y}{\partial z^2};$$

$$\mathrm{Re} = U_{\max} H / 2\nu; \qquad\qquad (18)$$

$$U_x(z) = 1 - z^2 : PP$$

$$U_x(z) = z : PC$$

The solution of equation (18) for perturbation of the velocity field must satisfy the same zero boundary conditions for PP flow and PC flow in the form:

$$u_y(x;\tau;z=\pm 1)=0 \qquad\qquad (19)$$

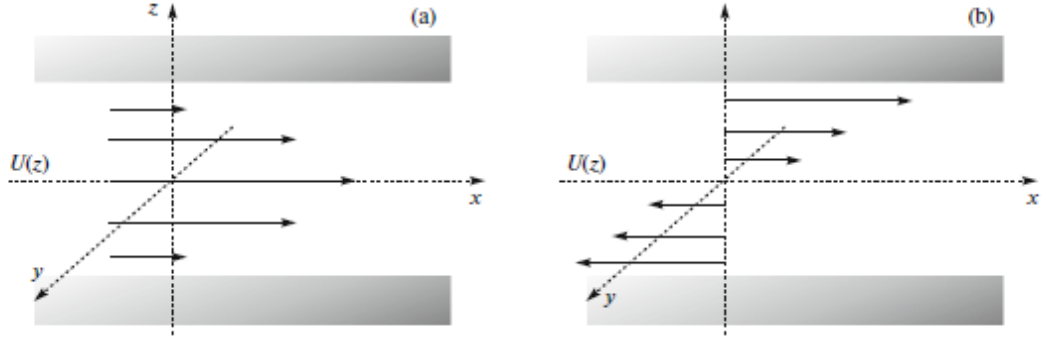

Fig. 1 Profiles of the main steady velocity $U(z)=U_x(z)$ normalized to $U_{\max}$ : a) for PP flow and b) for PC flow according to Eq. (18). Solid boundaries are represented here at $z=\pm 1$ in units of layer half-thickness $H/2$. (Fig. 1 is taken from Fig. 1 of [43])

In [43], the solution of equation (18) under boundary conditions (19) is sought in the form:

$$u_y=e^{S\tau}\sum_{n=1}^{N}\left[A_n(x)\sin(\pi nz)+B_n(x)\cos\left(\frac{\pi(2n-1)z}{2}\right)\right] \qquad (20)$$

Solution (20) satisfies the boundary conditions (19). In [43], conditions for linear instability of PP and PC flows were obtained in the case when, instead of the usual normal form of disturbances with the same period of longitudinal variability along the axis $x$, different periodic boundary conditions are used for each transverse modes $A_n;B_n$ with different numbers in the form:

$$\begin{aligned}
A_n(x)&=A_n(x+k_n^{-1});\\
B_n(x)&=B_n(x+k_n^{-1});\\
p_n&\equiv k_n/k_1\ne 1,2,3...;n=2,3,..
\end{aligned} \qquad (21)$$

It was shown in [43] that the linear instability of the PP flow with a positive exponential growth index $S>0$ of perturbations $u_y$ is possible (see Fig. 2a). For example, linear instability of PP flow arising for super threshold values of Reynolds numbers $\mathrm{Re}>\mathrm{Re}_{th}\approx 906$ in the case of two transverse modes $N=2$ in (20) with the values of $k_1\approx 0.675$ and $B_n=0$ , $p\approx 0.496$.

In the case $N = 100$ and $p \approx 0.5061$ in (20) the condition of linear instability $\mathrm{Re} > \mathrm{Re}_{th} \approx 1035$ obtained in [43] meets with the threshold value $\mathrm{Re}_{\exp} \approx 1080$ known from experimental data [3].

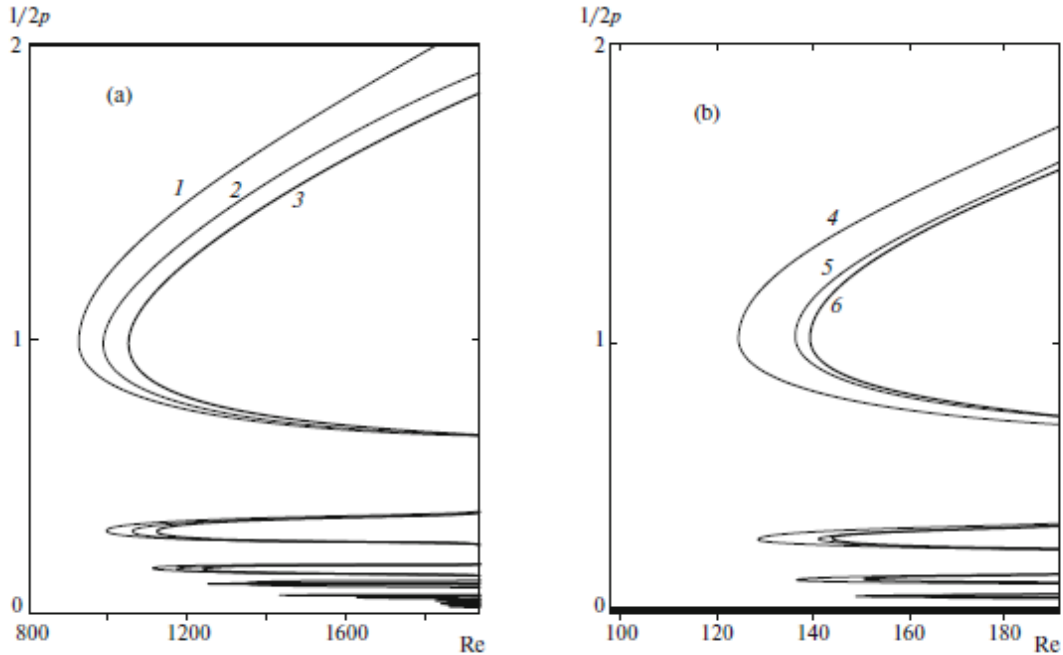

Fig. 2 Curves of neutral stability of the (a) PP flow for $k = 0.675$ and (b) PC flow for $k = 1.7037$: (1)- $\mathrm{Re}_{th\min} \approx 906.35$ for $N = 2; 1/2p = 1.008$ in (20) and (21); (2)- $\mathrm{Re}_{th\min} \approx 972.8$ for $N = 10; 1/2p = 0.988$; (3)- $\mathrm{Re}_{th\min} \approx 1035.31$ for $N = 100; 1/2p = 0.988$; (4)- $\mathrm{Re}_{th\min} \approx 124.27$ for $N = 2; 1/2p = 1.029$; (5)- $\mathrm{Re}_{th\min} \approx 136.475$ for $N = 10; 1/2p = 1.029$; (6)- $\mathrm{Re}_{th\min} \approx 139.077$ for $N = 100; 1/2p = 1.029$. (Fig. 2 is taken from Fig.2 [43]).

Also for the linear instability of the Couette flow in [43], in particular, when $A_n = B_n$ and $k_1 \approx 1.704$, the instability condition $\mathrm{Re} > \mathrm{Re}_{th} \approx 305$ is obtained for the ratio of periods $p \approx 0.746$, and for the value $p \approx 0.4859$ already at Reynolds numbers $\mathrm{Re} > \mathrm{Re}_{th} \approx 139$. This is in good agreement with experimental data [49]-[50] (see Fig. 3 below and Fig. 12 in [50]).

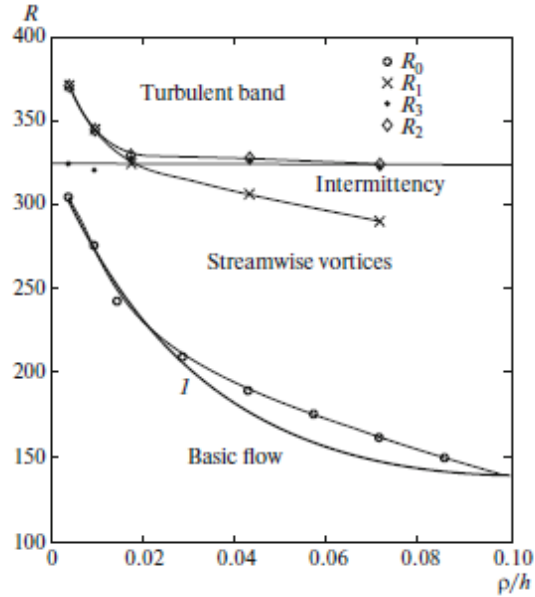

Fig. 3 Comparison theoretical curve 1, which is fragment of curve 6 from Fig. 2.b, with experimental data (see Fig. 12 in [50]) for PC flow instability; $1/2p$ value varies along the horizontal axis from 0.67 (corresponding to $\rho/h = 0.005$) to 1.029 (corresponding to $\rho/h = 0.1$).

In the case when the longitudinal periods of all transverse modes are equal in the limit $p \to 1$, then, as for the usual theory of linear stability of the PC flow, the Reynolds threshold number tends to infinity.

In this paper, by analogy with the consideration of the instability of the Hagen-Poiseuille flow, we show that even for the case of a single transverse mode N=1 in solution (20), it is possible to obtain conditions for linear instability of the PP and PC flows. This is indeed possible if, instead of the boundary conditions (21) of strict longitudinal periodicity of transverse modes, a quasi-periodic condition is used, similar to the boundary condition (8), which, when taking into account viscosity, must be considered instead of (21).

### 3.1 The plane Poiseuille (PP) flow

For example, for the case of PP flow, after substituting (20) into equation (18) in the case of a single nonzero transverse mode with $A_m \neq 0$, also after multiplying the result by $\sin \pi m z$ and averaging along the vertical coordinate from -1 to 1, we obtain the following equation for this mode (while the terms proportional to $B_m$ in (20) do not give a final contribution):

$$\frac{d^2 A_m}{dx^2} - \mathrm{Re}\left(\frac{2}{3} + \frac{1}{2\pi^2 m^2}\right)\frac{dA_m}{dx} - \left(S + \pi^2 m^2\right)A_m = 0 \qquad (22)$$

Note that equation (22) exactly coincides with equation (10) obtained in the previous section for the Hagen-Poiseuille flow if in (10) we replace

$P_{mm} \to \frac{2}{3} + \frac{1}{2\pi^2 m^2}$ and $j_{1,m}^2 \to \pi^2 m^2$.

As a result, based on the consideration of the solution of equation (22) under boundary conditions (8) and the use of a modification of representation (15) (in which it must be replaced $R \to H/2$), we obtain the following condition for linear instability of the plane Poiseuille flow:

$$\mathrm{Re} > \frac{6\pi^4 m^4}{(3 + 4\pi^2 m^2) k_1 \ln(1/\beta)};$$

$$k_1 \ln(1/\beta) = \frac{2\varepsilon}{H} \approx \frac{1}{\sqrt{2\,\mathrm{Re}}} \qquad (23)$$

$$\mathrm{Re} > \mathrm{Re}_{PP}(m=1) \approx 378.6; \quad \mathrm{Re} > \mathrm{Re}_{PP}(m=2) = 6754.3$$

### 3.2 The plane Couette (PC) flow

To study the linear stability of the Couette plane flow, we also use the representation of the solution in the form (20), and also only for the case of a single nonzero transverse mode.  The antisymmetry of the velocity profile of the stationary flow of the PC already leads to a finite contribution of the terms in (20), for which $B_m \neq 0$.

For simplicity, let us consider the case of identical constant coefficients $B_m = A_m$ in representation (20) to solve equation (18) describing the evolution of perturbations of the transverse component of the velocity field of a stationary Couette flow.

We substitute a solution of the form (20) into equation (18) in the case when it describes the evolution of a small perturbation of the transverse velocity component for a plane Couette flow in the case of a single nonzero mode with the number m. At the same time, before integrating over the variable $z$, we will also multiply by the function $\sin \pi m z$ of the result of substituting (20) into equation (18).

As a result, instead of equation (10) for the flow of HP or equation (23) for the flow of PP, we obtain a similar equation for the flow of PC in the form:

$$\frac{d^2 A_m}{dx^2} - \frac{\text{Re}}{\pi^2}\left(4 + \frac{1}{(2m-1/2)^2}\right)\frac{dA_m}{dx} - \left(S + \pi^2 m^2\right)A_m = 0 \qquad (24)$$

Accordingly, instead of (14) and (23), we obtain similar conditions for linear instability of the Couette plane flow in the form:

$$\text{Re} > \frac{\pi^4 m^2 (2m-1/2)^2}{\left(4(2m-1/2)^2+1\right)k_m \ln(1/\beta)};$$

$$k_m \ln(1/\beta) = \frac{2\varepsilon}{H} \approx \frac{1}{\sqrt{\text{Re}}}$$

$$\text{Re} > \text{Re}_{1PC}(m=1) \approx 480.4 \; ; \; \text{Re} > \text{Re}_{2PC}(m=2) \approx 56954.8 \qquad (25)$$

Note that the threshold Reynolds number in (25) for m=1 turns out to be comparable in magnitude to the threshold Reynolds number $\text{Re}_{TS} \approx 420$, which corresponds to observations of dissipative instability during the occurrence of Tollmien-Schlichting waves in the boundary layer [41]. This threshold number for the realization of linear instability also corresponds to the observed instability regime of the Couette plane flow, which leads directly to flow turbulence, according to Fig.3.

### 4. Landau's theory and dissipative instability of HP, PP, and PC flows

Let us consider the physical meaning of the conclusions obtained above about the linear instability of the HP, PP and PC flows, which is associated with the manifestation of instability caused by the action of a rather low, but non-zero viscosity.

Note that similar manifestations of dissipative instability, realized in various real physical systems [20]-[32], are associated with the energetically advantageous generation of perturbations [27], [22]. We will show the validity of this mechanism of dissipative instability in Section 4.1 using the example of the instability of the HP flow, and in sections 4.2 and 4.3 for the PP and PC flows, respectively. To do this, we use an approach similar to that developed in Landau's theory of superfluidity [22], which can equally be applied not only to the consideration of a quantum fluid, but also to a classical fluid in the HP, PP, and PC flows.

### 4.1 The Hagen-Poiseuille (HP)flow

Let us consider energetically the average flow velocity of HP described by equation (1), which is equal to $U_a = U_{\max} / \sqrt{3}$. This value, in contrast to the generally considered average cross-sectional velocity $\overline{U}_A = U_{\max} / 2$, is defined in terms of the average kinetic energy of fluid motion in HP flow (1):

$$E_A = MU_a^2 / 2 = \rho \int d^3x \overline{U}^2 / 2 = \rho \pi R^2 L U_{\max}^2 / 6 = MU_{\max}^2 / 6 \qquad (26)$$

In definition (26), $M = \rho \int d^3x = \rho \pi R^2 L$ -is the mass of a liquid in a certain sufficiently extended part of the flow of HP with some finite length $L$ and $\rho$ - is the constant density of an incompressible liquid.

Consider the Hagen-Poiseuille flow in a coordinate system that moves with a so-defined average velocity in the same direction as the velocity field (1):

$$U_a = U_{\max} / \sqrt{3} \qquad (27)$$

As in the work of Landau [22], we introduce an elementary vortex disturbance, which, if it occurs in a steady fluid, is characterized by energy $E_0(P_0)$ and momentum $\vec{P}_0$. Suppose that this small disturbance is also associated with a certain small mass $\Delta m \ll M$ of liquid, which is carried along by this vortex disturbance.

By analogy with the theory of Landau (1941) [22], we consider the following energy balance equation, which characterizes the possibility of generating an elementary disturbance with energy $E_1$ due to the interaction of the HP flow with the walls of the pipe, associated with the presence of a low, but not equal to zero shear viscosity:

$$MU_a^2 / 2 \geq E_1 + (M - \Delta m)U_a^2 / 2 \qquad (28)$$

If there is even a small viscous dissipation in (28), it is necessary to use the inequality sign. The ratio (28) for the value $\Delta m = 0$ exactly corresponds to the representation of the energy balance considered in [22]. According to [22], in order for the perturbation to be energetically beneficial, the left side of equality (28) must exceed the right side of equality. Therefore, the considered process of generating elementary excitation is an example of dissipative instability.

As in the Landau theory [22], in (28) we use the well-known relation [33] between the perturbation energy in a stationary medium and its representation in a coordinate system in which the liquid moves at a constant average velocity (27):

$$E_1 = E_0 + \left( \vec{P}_0 \vec{U}_a \right) = E_0 - P_{0z} U_a \qquad (29)$$

From the energy balance equation (28), taking into account the relation (29), we obtain a generalization of the well-known Landau criterion (1941) [22], which determines the condition for violation of the superfluid state due to the generation of elementary vortex excitations (rotons). This generalization to the case of the flow of a viscous incompressible fluid has the form:

$$U_a > \left( E_0 - \frac{\Delta m U_a^2}{2} \right) / P_{0z} \qquad (30)$$

For the value $\Delta m = 0$ condition (30) exactly coincides with the criterion for violation of superfluidity obtained in [22]. In the general case, from (30) we obtain the following generalization of the Landau criterion:

$$U_a > U_{th} = \frac{P_{0z}}{\Delta m} \left( \sqrt{1 + \frac{2 E_0 \Delta m}{P_{0z}^2}} - 1 \right) \qquad (31)$$

In the limit $2 E_0 \Delta m / P_{0z}^2 \ll 1$ from (31) follows the representation for the threshold velocity in the Landau criterion:

$$U_a > U_L = E_0 / P_{0z} \qquad (32)$$

In the opposite limit $2 E_0 \Delta m / P_{0z}^2 \gg 1$, an estimate of the threshold velocity follows from the generalization of the Landau criterion (31).

$$U_a > U_{NL} = \sqrt{\frac{2 E_0}{\Delta m}} \qquad (33)$$

Indeed, taking into account the finite mass value $\Delta m \neq 0$ in [61], a threshold condition of the form (33) is considered for the flow of a superfluid through a capillary when the value $U_a$ corresponds to the velocity of a superfluid quantum liquid. As a result, it was established in [61] that it is possible to eliminate the known [59] excess of the Landau superfluidity criterion [22] over the observed threshold velocities, precisely by taking into account the nonzero mass $\Delta m \neq 0$ of the corresponding elementary excitations. At the same time, it was found in [61] that the correspondence between the generalization of the Landau superfluidity criterion in the form of (33) and the experimental data [62] is obtained in the case when the mass of the resulting elementary excitation has a mass five orders of magnitude greater than the mass of helium atoms ($m_{_4He} \approx 6.65 \times 10^{-24} g$ -the mass of the helium atom). This is consistent with the well-known representation of Feynman [63] (see also [62], [64]-[68]) on the birth of extended Onsager-Feynman quantum vortices or closed vortex filaments (loops, rings) leading to the breakdown of superfluidity.

In this paper, based on the condition (31) of the threshold generation of disturbances for the flow of an ordinary classical (non-quantum) liquid, the possibility of disruption of the stability of the HP flow is also considered due to the mechanism of generation of vortex rings in the wall region of the flow due to the action of viscosity.

It is further shown that only in the limit of large Reynolds numbers does the condition of instability of the CP flow, obtained from the general condition (31), differ little quantitatively from condition (32), which coincides with the Landau criterion [22].

For the energy and momentum of a vortex ring, we will use the well-known representations [69] (see (7.2.16) and (7.2.14) in [69]):

$$E_0 \approx \frac{1}{2} \rho_0 R \kappa^2 \ln\left(\frac{R}{\varepsilon}\right) \qquad (34)$$

$$P_{0z} = \pi \rho_0 \kappa R^2 \qquad (35)$$

In (34) and (35) $R$ -is the radius of the vortex ring, which is approximately equal to the radius of the tube with a circular cross-section, through which the stationary flow of HP is carried out. In this case, the value of the circulation $\kappa$, and the radius of the vortex core $\varepsilon$, which depend on the viscosity and the Reynolds number, for example, according to the estimate {57}:

$$\varepsilon / R \approx p / \mathrm{Re}^m ;$$
$$1/2 \geq m \geq 1/4 \tag{36}$$

For simplicity, we first estimate the threshold Reynolds number based on condition (32), which, in the case of the threshold generation of disturbances in the form of vortex rings, taking into account relations (34) and (35), has the form:

$$U_a > 2U_V = \frac{\kappa}{2\pi R} \ln\left(\frac{R}{\varepsilon}\right) \tag{37}$$

It follows from condition (37) of instability of the HP flow that the generation of a disturbance in the form of a vortex ring is energetically advantageous only at the threshold value of the average HP flow velocity, which is equal to twice the velocity of the vortex ring (see (7.2.15) in [69]):

$$U_V = \frac{\kappa}{4\pi R} \ln\left(\frac{R}{\varepsilon}\right) \tag{38}$$

In (38) $U_V$ - is the value of the velocity of the vortex ring in an unlimited space, with which the center of gravity of the vortex ring moves due to the self-interaction.

From the instability criterion (32), taking into account the determination of the average velocity (27), we obtain the following conditions for instability of the HP flow:

$$\mathrm{Re} > \frac{\sqrt{3}}{2\pi} \frac{\kappa}{\nu} \ln\left(\frac{R}{\varepsilon}\right) \tag{39}$$

Let us consider in (39) the following estimate of the circulation value $\kappa$ as a function of the radius of the vortex core $\varepsilon$:

$$\kappa / \nu \cong 2\pi\varepsilon U_\varepsilon / \nu \approx \frac{8\pi\varepsilon^2}{R^2} \mathrm{Re} \approx 8\pi (20\mathrm{Re})^{1/2} \tag{40}$$

When obtaining the estimate (40), an estimate of the value $U_\varepsilon = U_{\max}\left(1 - (R - 2\varepsilon)^2 / R^2\right) \approx 4\varepsilon U_{\max} / R$ of the flow velocity at a distance $2\varepsilon$ from the pipe wall is used.

Using representation (36) for the value $\varepsilon/R$ in (39) and (40) at a value $p = \gamma/q \approx 2.1147425$; $m = 1/4$ in (36) (see Fig. 10 and Fig.11 in [57], where $\gamma \equiv 2\pi\varepsilon/\lambda = 1.4$ and $2\pi R/\lambda = n \approx q\,\mathrm{Re}^m$ for $q \approx 0.662$), we obtain the following condition for instability of the laminar flow of HP due to the energetically favorable spontaneous generation of the vortex ring:

$$\mathrm{Re} > \mathrm{Re}_{th} \approx 837 \qquad (41)$$

Indeed, according to (36), for values of parameters $m = 1/4$; $p = 20^{1/4} \approx 2.1147$ satisfying the observational data of vortex rings [57] from (39) and (40), we obtain an inequality $\mathrm{Re} > 60\ln^2(\mathrm{Re}/20)$, from which the condition for instability of the HP flow in the form (41) is stated.

Now we obtain the threshold value of the Reynolds number based on the initial formula (31). To do this, we define in the following form the value of the mass of the liquid involved in the motion of the vortex ring, the creation of which leads to the dissipative instability of the laminar flow of HP

$$\Delta m = \rho_0 2\pi^2 R\varepsilon^2 \qquad (44)$$

From (44) and (35), taking into account (40) and (27), we obtain a relation $P_{0z}/\Delta m = 4U_{\max}$ that allows us to reduce the condition of instability of the flow of HP (31) to the inequality:

$$1 > 6.928\left(\sqrt{1 + \frac{2\varepsilon^2}{R^2}\ln\left(\frac{R}{\varepsilon}\right)} - 1\right) \qquad (45)$$

Condition (45), taking into account (40), leads to the condition of instability of the HP flow in the form:

$$\mathrm{Re} > \mathrm{Re}_{th} \approx 610 \qquad (46)$$

Thus, the estimate (46) differs significantly from the estimate of the threshold Reynolds number in (41), which is obtained based on formula (32), which follows from (31) only in the limit of large Reynolds numbers.

We also note that the threshold value of the Reynolds number obtained in (46) is in good agreement with the estimate of the minimum Reynolds number obtained in [42], at which the HP flow becomes unstable when using one hundred radial modes in the representation for the field of disturbances (3)-(5) (N=100 in (5)).

Thus, for the values of the Reynolds numbers under consideration, it is more useful to use a generalization of the Landau criterion in the form of (31), which takes into account the finite nonzero mass of the liquid entrained by the vortex ring field, the occurrence of which leads to instability of the HP flow.

## 4.2 The plane Poiseuille flow

To obtain a generalization of Landau's theory for the case of PP flow in a classical viscous incompressible fluid, we introduce, by analogy with (26) and (27), the energetically average PP flow velocity in the following form:

$$U_a = U_{max}\sqrt{2/3} \qquad (47)$$

As for the HP flow, the value of the average PP flow velocity is determined not as the average over the flow section, but as (26) for the PP flow velocity distribution, represented in dimensionless form in (18).

As a result, for the threshold conditions for the occurrence of dissipative instability of the PP flow, due to the mechanism of energetically favorable generation of vortex disturbances near solid boundaries, it has the same form (31) and (32) as for the HP flow.

However, due to the difference between the symmetry of the PP flow and the symmetry of the HP flow, the type of vortex disturbances will no longer be associated with the spontaneous formation of vortex rings when conditions (31) or (32) are met. Instead of vortex rings in the coordinate system moving with the average velocity (47) of the PP flow, a homogeneous and extended vortex sheet along the y- axis should be created near each of the two solid boundaries limiting the PP flow. In this case, the vortex sheet at any of the boundaries should have a circulation equal in magnitude, but opposite in sign to the magnitude of the circulation of the velocity field of the vortex sheet, which occurs simultaneously at the opposite solid boundary with a finite nonzero shear viscosity. Therefore, a system of these two vortex sheets can be modeled using a pair of point vortices on a plane transverse to the direction of the PP flow. Due to the viscosity, each of these vortices has a finite vortex core of radius $\varepsilon$, defined as (23).

As a result, under conditions of dissipative instability (31) or (32), instead of representations (34), (35) for the case of PP flow we will use the energy and momentum of a pair of point vortices on a plane (for a layer of unit thickness) in the form (see (7.3.7), (7.3.8), (7.3.14) in [69]):

$$|E_0| = \frac{\rho_0}{2\pi} \kappa^2 \ln\left(\frac{H}{\varepsilon}\right) \qquad (48)$$

$$P_{0x} = \rho_0 \kappa H \qquad (49)$$

In (48), (49), the magnitude of the energy and momentum of the vortex pair refers to a layer of liquid of unit thickness in the direction of the coordinate axis coinciding with the direction of the flow of PP, that is, the direction of the $x$-axis.

After substituting (48) and (49) in (32), we obtain the condition for instability of the PP flow in a form similar to condition (39):

$$\text{Re} > \frac{\sqrt{3/2}}{4\pi} \frac{\kappa}{\nu} \ln\left(\frac{H}{\varepsilon}\right) \qquad (50)$$

When taking into account the modification of ratios (36) and (40), which take into account the specifics of the PP flow, we will assume that the ratios are fulfilled in the form:

$$2\varepsilon / H \approx 2.828 / \text{Re}^{1/4}; \qquad (51)$$

$$\kappa / \nu \cong 2\pi\varepsilon U_\varepsilon / \nu \approx \frac{32\pi\varepsilon^2}{H^2} \text{Re} \approx 64\pi \, \text{Re}^{1/2} \qquad (52)$$

From (50)-(52) we obtain an inequality $\text{Re} > 24 \ln^2(\text{Re}/4)$ that holds under the condition:

$$\text{Re} > \text{Re}_{PP} \approx 605 \qquad (53)$$

Taking into account the finiteness of the mass $\Delta m = \rho_0 \pi \varepsilon^2$ per unit length of the vortex sheet occurs at the above-threshold average flow velocity PP (31) near the solid boundary due to the viscosity of (31), (47)-(49), (51), (52) we obtain the condition of instability of the PP flow in the form of inequality

$$1 > 19.595917\left(\sqrt{1 + \frac{1}{2\sqrt{\text{Re}}} \ln(\text{Re}/4)} - 1\right).$$

This inequality holds for above-threshold Reynolds numbers.:

$$\mathrm{Re} > \mathrm{Re}_{PP} \approx 556 \qquad (54)$$

As in the case of HP flow, the value of the Reynolds threshold number (54) turns out to be noticeably lower than the estimate (53) obtained on the basis of the well-known form of the Landau criterion (32), which does not take into account the finiteness of the mass of the liquid entrained by elementary vortex excitation.

### 4.3 The plane Couette flow

For the plane Couette current (PC), the average cross-sectional flow velocity is zero, unlike the HP and PP currents. However, the average velocity of the PC flow, determined by analogy with (26) and characterizing the average kinetic energy of some characteristic finite part of the PC flow, is finite and has the same form (27), as for the average velocity of the HP flow.

Let us consider a generalization of the Landau criterion presented in (31) and (32) for the PC flow describing the stationary laminar motion of a classical viscous incompressible fluid. As in the cases of the HP and PP flows, we determine the threshold of instability of the PC flow relative to the spontaneous energetically favorable appearance of a vortex shroud near one of the moving solid boundaries of the PC flow. Indeed, for the PC flow, due to the asymmetry of the flow, a vortex disturbance occurs only near that solid boundary, the direction of movement of which coincides with the direction of the average PC flow velocity. In (27), only the value of the average flow velocity of the PC is determined, and the direction can be chosen along both the positive and negative directions of the axis x. This distinguishes the PC flow from the HP and PP flows, where the average flow velocity has a direction that coincides with the average velocity over cross-section for these flows.

For example, if the average velocity of PC flow is directed along the positive direction of x- axis, then in the coordinate system moving with an average flow velocity (27), the fluid has zero velocity in a plane with a coordinate $z = z_0 = H/2\sqrt{3}$ at a distance $\lambda_0 = \dfrac{H}{2} - z_0 \approx 0.4226 H / 2$ from the upper boundary. In this case, it is near the upper boundary that a vortex sheet may occur when condition (31) or (32) is met, and this vortex sheet can be modeled using a single point vortex on the plane. This vortex should be located near the upper boundary at a distance of

$$\varepsilon = \frac{H}{2} p \, \mathrm{Re}^{-m} \le l \le \lambda_0 \qquad (55)$$

Due to the interaction with its mirror image, this point vortex will move in the same direction as the upper boundary with a velocity скоростью $U_V = \kappa / 4\pi l$ [69] (see Fig. 7.3.1 in [69]).

A vortex pair formed by a vortex and its mirror image has energy and momentum [69]:

$$|E_0| = \frac{\rho_0}{2\pi} \kappa^2 \ln\left(\frac{2l}{\varepsilon}\right) \qquad (56)$$

$$P_{0x} = 2\rho_0 \kappa l$$

From (27), (32) and (56) we obtain the instability condition of the PC flow in the form:

$$\mathrm{Re} > \frac{\sqrt{3}}{8\pi} \frac{H}{l} \frac{\kappa}{\nu} \ln\left(\frac{2l}{\varepsilon}\right) \qquad (57)$$

In (57) we use a representation similar to (52) having the form:

$$\frac{\kappa}{\nu} = \frac{4\pi\varepsilon^2}{H^2} \mathrm{Re} \qquad (58)$$

An estimate $\kappa \cong 2U_{\max}\pi\varepsilon^2 / H$ is used to obtain (58).

From (57), taking into account (58) and (55) in the case $m = 1/4$ and $p = 14.4$, we obtain the condition of instability of the PC flow:

$$\mathrm{Re} > 350 \qquad (59)$$

This value is consistent with the estimate of the Reynolds threshold number obtained in [43] and the observational data shown in Fig.3 with a parameter value of 0.01 along the abscissa axis in Fig.3 (turbulence region).

**5. Discussion**

The Introduction notes the dominant approach so far of the need to circumvent the problem of linear stability of the Hagen-Poiseuille (CP) and plane Couette (PC) flows by involving the idea that the instability of these flows observed in the experiment is due to the nonlinear effect of the finiteness of the amplitude of disturbances. At the same time, a fundamental question still remains unanswered: "why does the addition of viscous forces to an otherwise stable shear flow sometimes render it unstable" [59], несмотря на многочисленные попытки найти этот ответ [70]-[72] (см. также ссылки в [59] and [72]).

However, this issue does not yet relate to the problem of linear stability of HP and PC (with the exception of its consideration in our works [40]-[43]), but is mentioned only in connection with the established linear instability of the plane Poiseuille flow [17], [73] and the laminar stationary Blasius flow [59], [70]-[72]. The dissipative instability observed in the form of Tollmien-Schlichting waves [58] in the boundary layer was noted above. It is implemented for the Blasius flow, which is stable at zero viscosity and has the form:

$$U_x(z) = U_\infty \left( 1 - \exp\left( -\frac{z}{\delta} \right) \right) \qquad (60)$$

particular, it is noted in [59]: "This is one of the classical paradoxes of fluid mechanics". Lighthill (1963), Lindzen (1988) and others has proposed solution to it, but no general consensus as to a clear physical picture has yet emerged [59].

In our work, we propose a physical mechanism for the instability of HP, PP, and PC flows, which is caused by the action of viscosity in explicit form, which leads to the energetically advantageous generation of vortex disturbances near solid surfaces that limit these flows of a viscous incompressible fluid. The rationale for this mechanism was carried out above, both by obtaining a generalization of Landau theory in section 4 [22], and by rejecting in sections 2 and 3 the traditionally considered assumption in the theory of hydrodynamic stability about the need to use only strictly periodic disturbances along the direction of the main flow.

It is shown that if any viscosity other than zero is taken into account, it is the quasi-periodic boundary conditions (8) used earlier in connection with the study of the stability of the shock wave front by Brushlinsky [55] that need to be introduced into consideration. It is precisely this replacement of traditional periodic boundary conditions with quasi-periodic boundary conditions determined by the characteristic size of small vortex disturbances that makes it possible to eliminate this age-old paradox in hydrodynamics.

It should be noted that the idea of Dan Shechtman (1988) [74], which similarly shook the unshakable idea of the need to consider only strictly periodic spatial structures in crystallography, was not immediately accepted, but nevertheless received an assessment by the Nobel Committee in 2011. Therefore, this should more naturally be accepted in hydrodynamics, where the understanding of the deliberate idealization of the assumption of strict periodicity of disturbances has long been well understood [19].

We also note some similarity between the theory [40]-[43] and the approaches proposed in our present work with the mechanism of dissipative instability of the Blasius flow (60), considered in [59], manifested in the form of the generation of Tolmin-Schlichting waves [58].

In [59], although within the framework of the traditional consideration of only strictly periodic disturbances in the direction of the propagation of the Blasius flow (60), it is the mechanism of dissipative instability of this flow that is proposed, modeled as (see (2.9) in [59]):

$$U_x(z) = U_0 \frac{z}{d}, 0 \le z \le d - \delta; \qquad (61)$$

$$U_x(z) = U_0 \left( 1 - \frac{(d + \delta - z)^2}{4d\delta} \right), d - \delta \le z \le d + \delta \qquad (62)$$

$$U_x(z) = U_0, z > d + \delta \qquad (63)$$

It was noted in [59] that the set of representations (61)-(63) approximately describes the Blasius flow (60) at the value $\delta/d \approx 0.36$. In [59], the instability of the flow (61)-(63) is investigated with respect to two-dimensional small perturbations of the velocity field, when, in contrast to the consideration based on equation (18): $u_x \ne 0; u_z \ne 0; u_y = 0$ .

In [59], the perturbation is sought in the form of a traveling wave, when the consideration is carried out in a coordinate system that moves with a constant velocity of this wave. In this moving coordinate system, the velocity of the main perturbed flow (60) or (61)-(63) already becomes nonzero at the solid boundary of the flow at $z = 0$ and having a direction opposite to the initial main flow near $z = 0$. This is qualitatively consistent with our consideration of the instability of the Couette flow in section 4.3 when generalizing the Landau criterion. Moreover, representation (61) corresponds to the Couette flow velocity profile. In [59], this change in the direction of velocity near the solid boundary is used to introduce an additional viscous mode, which must be excited due to the need to meet the non-slip condition for it. This viscous mode, although decaying in time, interacts with the initial inviscid mode and can enhance it for some sufficiently large Reynolds numbers (see Fig. 7 in [59]). In [59], a representation similar to (55) is used to describe the characteristic size of vortices generated near the boundary in the case $m = 1/3$.

Thus, the mechanism of dissipative instability of the Blasius flow proposed in [59] is due to the need for resonant interaction of viscous and inviscid modes. It can correspond to both the generalization of the Landau criterion discussed in section 4.3 and the linear interaction of two or more radial modes having different periods of longitudinal variability in theory [40]-[43].

But the approach developed in sections 2 and 3, based on consideration of quasi-periodic boundary conditions due to finite viscosity, provides a relatively simple approach and clear understanding of the mechanism of linear instability of shear flows. This approach can be applied to the Blasius flow (60) if equation (18) uses the representation for the velocity field of the main flow in the form (60). Similarly, a generalization of Landau theory can be used for the Blasius flow if the average flow velocity for the flow (60) is determined energetically based on an appropriate modification of the condition (26).

## Conclusions

Thus we show the possibility of realizing the linear instability of HP, PP, and PC flows due to the destabilizing effect of the low but nonzero viscosity. Indeed, for no-zero viscosity the quasi-periodic boundary conditions is need to taken into account instead of the traditional used idealized exact periodic boundary condition. A generalization of the Landau theory has been carried out, as a result of which the conditions of instability of the HP, PP and PC flows have been established. That instability of different bounded shear flows are also determined by a mechanism related to the effect of viscosity in the energetically favorable arising of vortex disturbances.